\documentclass{emulateapj}
\usepackage{graphicx}
\usepackage{txfonts}
\usepackage{natbib}
\usepackage{longtable} 

\makeatletter
{\end{center}\end{minipage}\smallskip}

\def\ltsima{$\; \buildrel < \over \sim \;$}
\def\lsim{\lower.5ex\hbox{\ltsima}}
\def\loe{\lower.5ex\hbox{\ltsima}}
\def\gtsima{$\; \buildrel > \over \sim \;$}
\def\gsim{\lower.5ex\hbox{\gtsima}}
\def\goe{\lower.5ex\hbox{\gtsima}}
\def\ergs{\rm \ erg \, s^{-1}}

\def\ltsima{$\; \buildrel < \over \sim \;$}
\def\lsim{\lower.5ex\hbox{\ltsima}}
\def\loe{\lower.5ex\hbox{\ltsima}}
\def\gtsima{$\; \buildrel > \over \sim \;$}
\def\gsim{\lower.5ex\hbox{\gtsima}}
\def\goe{\lower.5ex\hbox{\gtsima}}

\def\sgra{\mbox{SGR\,1806--20}}

\def\rxj {1RXS J170849$-$400910}
\def\ea {1E\,2259$+$586}
\def\src{SGR\,1900$+$14}

\def\cxo{CXOU J164710.2$-$455216}
\def\swift{{\em Swift}}
\def\ergscm2{\rm erg\,cm^{-2}\,s^{-1}}
\def\ergs{\rm erg\,s^{-1}}

\def\XMM{{\em XMM-Newton}}

\lefthead{ISRAEL ET AL.}
\righthead{A SWIFT GAZE INTO THE 2006 FOREST OF SGR\,1900+14}
\submitted{Received:  21 December 2007; ~~ Accepted: 23 May 2008}

\begin{document}

\title{A \swift\ gaze into the 2006 March 29$^{th}$ burst forest of
SGR\,1900$+$14}

\author{G.L.\ Israel\altaffilmark{1}, P.\ Romano\altaffilmark{2,3,4}, 
V.\ Mangano\altaffilmark{4}, S.\ Dall'Osso\altaffilmark{5,1}, 
G.\ Chincarini\altaffilmark{2,3}, L.\ Stella\altaffilmark{1},
S.\ Campana\altaffilmark{2}, T.\ Belloni\altaffilmark{2}, G.
Tagliaferri\altaffilmark{2}, A.J.\ Blustin\altaffilmark{6}, 
T.\ Sakamoto\altaffilmark{7}, K.\ Hurley\altaffilmark{8},
S.\ Zane\altaffilmark{6}, A.\ Moretti\altaffilmark{2}, 
D.\ Palmer\altaffilmark{9}, C.\ Guidorzi\altaffilmark{2,3}, 
D.N.\ Burrows\altaffilmark{10}, N.\ Gehrels\altaffilmark{7}, 
H.A.\ Krimm\altaffilmark{7,11}}

\affil{1. INAF -- Osservatorio Astronomico di Roma, Via Frascati 33, 
       I--00040 Monteporzio Catone (Roma),  Italy; 
       gianluca, stella@mporzio.astro.it}

\affil{2. INAF -- Osservatorio Astronomico di Brera, Via Bianchi
	46, I--23807 Merate (Lc), Italy}

\affil{3. Universit\`a{} degli Studi di Milano, Bicocca, Piazza 
delle Scienze 3, I-20126, Milano, Italy}

\affil{4. INAF -- Istituto di Astrofisica Spaziale 
                 e Fisica Cosmica Sezione di Palermo, 
                 via Ugo La Malfa 153, I-90146 Palermo, Italy}

\affil{5. Dipartmento di Fisica ``Enrico Fermi'', Universit\'a di Pisa,
Largo B. Pontecorvo 3, I-56127 Pisa, Italy}

\affil{6. Mullard Space Science Laboratory, University College London,
Holmbury St. Mary, Dorking Surrey, RH5 6NT, UK}	

\affil{7. NASA/Goddard Space Flight Center, Greenbelt, MD 20771, USA}

\affil{8. Space Sciences Laboratory, University of California at
Berkeley, 7 Gauss Way, Berkeley, California, 94720-7450, USA}

\affil{9. Los Alamos National Laboratory, Los Alamos, NM, 87545, USA}

\affil{10.  Department of Astronomy \& Astrophysics, Pennsylvania State
University, University Park, PA 16802, USA}

\affil{11. Universities Space Research Association, 10211
Wincopin Circle, Suite 500, Columbia, MD 21044}

\begin{abstract}
In 2006 March the Soft Gamma-ray Repeater \src\ resumed its bursting
activity after about 2 years of quiescence.  The \swift\ mission observed the
source several times in order to monitor its timing and spectral properties.  We
report on the intense burst ``forest'' recorded on 2006 March 29
which lasted for $\sim$30\,s, when Swift was pointing at the source with
the narrow field of view instruments. More than 40 bursts were detected both by 
BAT and by XRT, seven of which are rare intermediate flares (IFs): several times
10$^{42}$\,ergs were released. The BAT data were used to carry out time-resolved
spectroscopy in the 14--100\,keV range down to 8\,ms timescales. BAT and XRT
simultaneous data (in the 1--100\,keV range) were used to characterize the
broad-band energy spectra of IFs and verify the results obtained from the
BAT-only spectral fits.

This unique dataset allowed us to test the magnetar model
predictions such as the magnetically trapped fireball and the
twisted magnetosphere over an unprecedented range of fluxes and with large
statistics (in terms of both photons and IFs). We confirmed that a two blackbody
component fits adequately the time-resolved and integrated spectra of IFs.
However, Comptonization models give comparable good
reduced $\chi^2$. Moreover, we found: i) a change of behavior, around $\sim
10^{41}\,\ergs$, above which the softer blackbody shows a sort of saturation
while the harder one still grows to a few times $10^{41}\ergs$; ii)  a rather
sharp correlation between temperature and radii of the blackbodies ($R^2 \propto
kT^{-3}$), which holds for the most luminous parts of the flares (approximately
for $L_{\rm tot}\geq 10^{41}\,\ergs$).
Within the magnetar model, the majority of these findings can be accounted
for in terms of thermalised emission from the E-mode and O-mode
photospheres. Interestingly, the maximum observed luminosity coming from a
region of $\sim15$\,km matches the magnetic Eddington luminosity at the same
radius, for a surface dipole field of $\sim 8 \times 10^{14}$\,G (virtually 
equal to the one deduced from the spindown of \src). 

We also studied the persistent emission of \src\ preceding and following the
burst ``forest'' in search for transient and permanent timing and spectral
variations on short (days/week) timescales. We found
that the timing/spectral property variations are correlated with the flux and
are likely only due to changes of the high energy part of the spectrum (above
5\,keV), further supporting the twisted ``magnetosphere'' scenario.  
\end{abstract}


\section{Introduction}
\label{sec:introd}
Soft Gamma-ray Repeaters (SGRs) are a small class of high energy transient
astrophysical  sources characterized by their emission of short ($\sim100$\,ms)
bright ($10^{39}$--$10^{42}\,\ergs$)  bursts of soft  $\gamma$ rays. 
SGRs have been associated with persistent X--ray counterparts with 1--10\,keV 
luminosity of $\sim 10^{34}$--$10^{35}\,\ergs$ and energy spectra that are
generally well fit with a power-law model ($\Gamma \sim$2). A blackbody
component with $kT \sim$0.5\,keV has been found in \src\ and \sgra\ (Woods et
al.\ 1999; Woods et al.\ 2004; Mereghetti et al.\ 2005). 
In contrast to the common short bursts, three SGRs have emitted one
very powerful long-duration burst. 
These ``giant flares'' are distinguished by their extreme energies
($10^{44}$--$10^{46.5}\,\ergs$), their hard spectra at the onset, and the
presence of coherent 5-7\,s pulsations at the spin period of the neutron star
during the decaying tail lasting several minutes.
\begin{deluxetable*}{lllll}
  \tablewidth{0pc} 	      	
  \tabletypesize{\scriptsize} 
  \tablecaption{\src\ Swift observation log for pointings
carried out during 2006.\label{sgr:alldata}} 
  \tablehead{
\colhead{Sequence} &    \colhead{Obs/Mode} &    \colhead{Start time  (UT)} &   
\colhead{End time (UT)} & \colhead{Exposure}  \\
\colhead{} &  \colhead{} &    \colhead{(yyyy-mm-dd hh:mm:ss)} &
\colhead{(yyyy-mm-dd hh:mm:ss)} &    \colhead{(s)} 
}
 \startdata	
00202746000\tablenotemark{a} &       BAT/EVENT &       2006-03-25 20:16:30     &
      2006-03-25 20:17:13     &       43      \\
00203045000\tablenotemark{b} &       BAT/EVENT &       2006-03-28 13:50:05     &
      2006-03-28 13:50:48     &       43      \\
00203109000\tablenotemark{c} &       BAT/EVENT &       2006-03-29 01:27:53     &
      2006-03-29 01:28:36     &       43      \\
00203127000\tablenotemark{d} &       BAT/EVENT &       2006-03-29 02:52:50     &
      2006-03-29 02:53:42     &       50      \\
00203974000\tablenotemark{e} &       BAT/EVENT &       2006-04-05 06:43:21     &
      2006-04-05 06:44:04     &       43      \\
00214277000\tablenotemark{f} & BAT/EVENT & 2006-06-10 06:49:02 & 2006-06-10
07:09:04 & 1202 \\
\hline
00030386001	&	XRT/PC	&	2006-03-25 22:50:43	&
2006-03-27 21:47:47	&	49564	\\
00030386002	&	XRT/PC	&	2006-03-28 01:05:28	&
2006-03-28 23:48:57	&	11350	\\
00030386003	&	XRT/WT 	&	2006-03-29 01:01:00	&
2006-03-29 23:57:00	&	18592	\\
00030386004	&	XRT/WT 	&	2006-03-30 01:07:00	&
2006-03-30 23:59:58	&	20055	\\
00030386005	&	XRT/WT 	&	2006-03-31 01:26:23	&
2006-03-31 23:59:59	&	15931	\\
00030386006	&	XRT/WT 	&	2006-04-01 01:13:58	&
2006-04-03 13:12:31	&	45951	\\
00030386008	&	XRT/WT 	&	2006-04-07 11:48:02	&
2006-04-07 23:20:00	&	7846	\\
00030386009	&	XRT/PC	&	2006-04-08 00:51:04	&
2006-04-10 23:38:58	&	29274	\\
00030386010	&	XRT/PC	&	2006-04-11 00:48:12	&
2006-04-11 23:44:55	&	18250	\\
00030386011	&	XRT/PC	&	2006-04-12 00:47:49	&
2006-04-12 23:43:57	&	20503	\\
00030386012	&	XRT/PC	&	2006-04-13 00:53:42	&
2006-04-13 18:41:34	&	10702	\\
00030386013	&	XRT/PC	&	2006-04-14 22:22:41	&
2006-04-14 23:59:56	&	1602   \\
00214277000     &       XRT/PC  &       2006-06-10 06:54:59     &      
2006-06-10 08:56:03     &       4071
    \enddata 
    \tablenotetext{a}{First BAT trigger: 2006-03-25   20:16:40.28 UT.} 
    \tablenotetext{b}{Second BAT trigger: 2006-03-28   13:50:15.00 UT.} 
     \tablenotetext{c}{Third BAT trigger: 2006-03-29   01:28:03.99 UT.} 
     \tablenotetext{d}{Fourth BAT trigger: 2006-03-29   02:53:09.46 UT.} 
    \tablenotetext{e}{Fifth BAT trigger:   2006-04-05  06:43:31.36 UT.} 
    \tablenotetext{f}{Sixth BAT trigger:   2006-06-10   06:53:00.8 UT.}
   \end{deluxetable*}  

Both \sgra\ and \src\ were found to spin down secularly at  a rate of
$\sim10^{-11}$ to $10^{-10}$ s s$^{-1}$ (Kouveliotou et al.\ 1998,1999).
Combining SGR periods and their first derivatives, and assuming that
the $B_{\rm dip}\sim3.2 \times 10^{19} (P \dot{P})^{1/2}$\,G relationship
generally applied to radio pulsars holds for SGRs as well, magnetic field
strengths of $10^{14}$--$10^{15}$\,G are inferred. The latter values provide
strong evidence that SGRs are highly magnetized neutron stars, i.e.,
``magnetars'' (Duncan \& Thompson 1992). In the  magnetar model, the magnetic
field is the dominant source of free energy (orders of magnitudes higher
than the rotational energy of the star), powering the persistent emission
through low-level seismic activity and heating of the stellar interior (Thompson
\& Duncan 1996). When magnetic stresses build up sufficiently to crack a patch
of the neutron star crust, the resulting ``crustquake'' ejects hot plasma
particles into the magnetosphere, which results in
an SGR burst (Thompson \& Duncan 1995). Alternatively, the short SGR bursts may
arise from magnetic reconnection events in the stellar magnetosphere
(Lyutikov 2002). Giant flares likely result from a sudden  reconfiguration
of the star's magnetic field that produces large fractures in the crust and
propagates outwards through Alfv\'en waves of enormous power. The
$\sim$0.2\,s initial spike of giant flares marks the ejection of the largest
part of the total event energy (up to $\sim$10$^{47}$\,ergs) as well as the
onset of a relatively large surface fracture (Schwartz et al. 2005). The
detection of deca-, hecto-, and kilo-Hertz QPOs in the decaying phases (tails)
of giant flares have been interpreted in terms of global seismic oscillations
following the fracture formation (Israel et al. 2005; Strohmayer \& Watts
2005).The super-Eddington fluxes observed in the tails of these events are
thought to be possible because of the suppression of the electron scattering
cross sections in the presence of very strong magnetic fields (Paczy\'nski
1992). 

Four confirmed SGRs  are known so far, three of which are in our Galaxy and
one in the Large Magellanic Cloud. The Galactic SGRs are located very close to
the plane of the Galaxy, indicating that SGRs belong to a young stellar
population.
Furthermore, two of these SGRs are likely associated with clusters of very
massive stars (Fuchs et al.\ 1999; Vrba et al.\ 2000; Eikenberry et al.\ 2001).
\begin{figure*}
\begin{center}
\epsscale{.80}
\includegraphics[angle=-90,scale=.65]{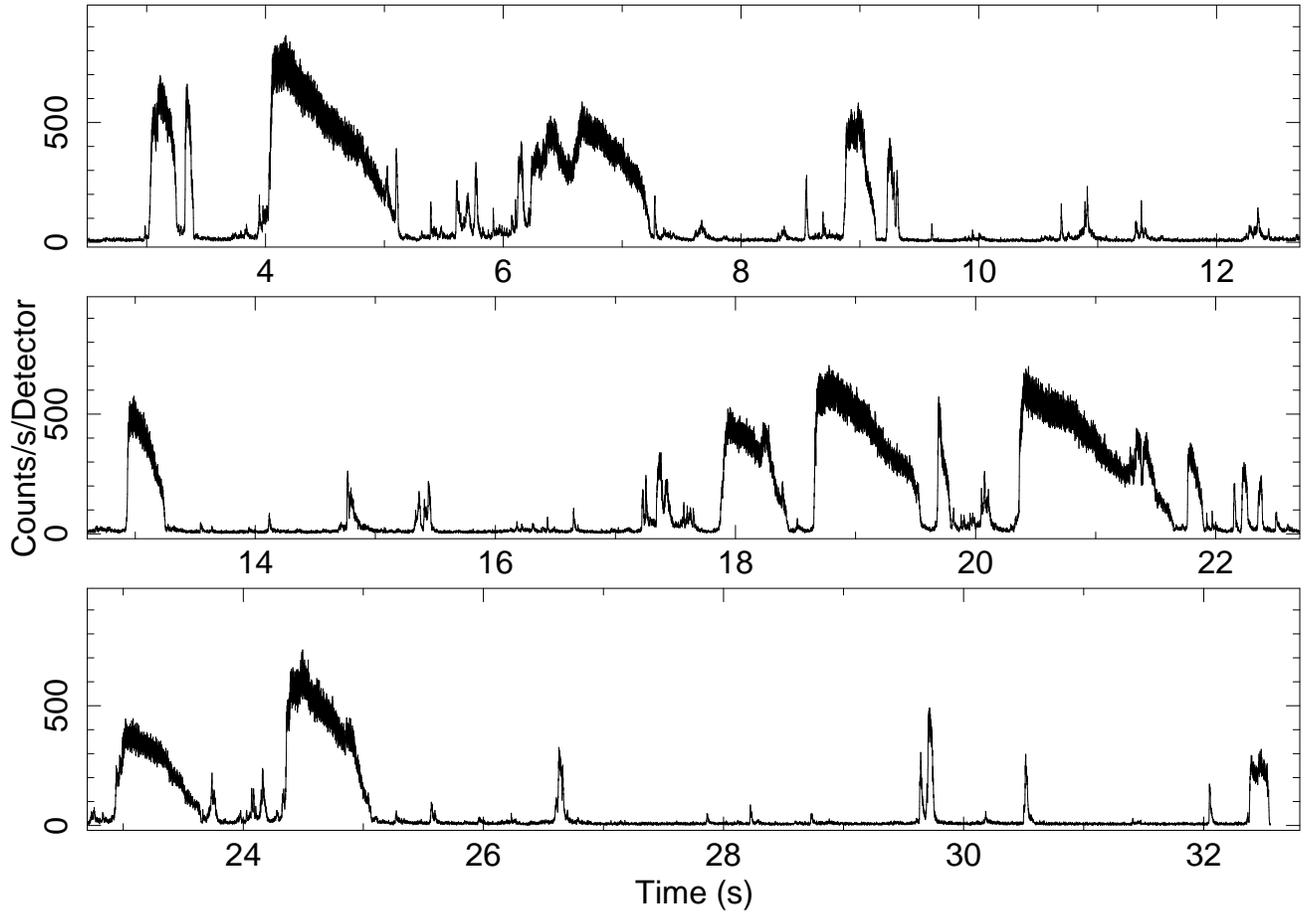}
\caption{15--100keV BAT light curves with a time resolution of 1\,ms obtained 
during the burst ``forest'' of 2006 March 29.  
}
\label{fig:BATlcurve}
 
\end{center}
\end{figure*}

\begin{figure}
\begin{center}
\epsscale{.90}
\includegraphics[angle=-90,scale=.55]{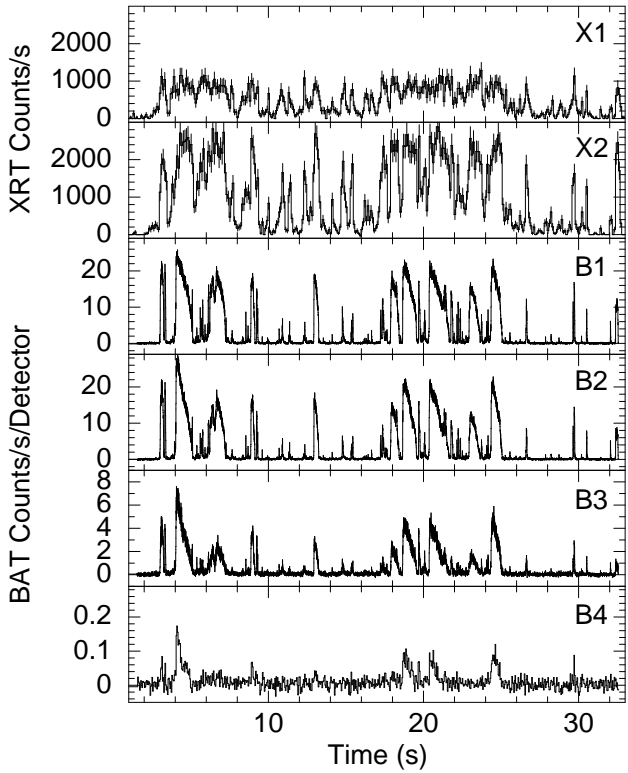}
\caption{BAT and XRT (WT) light curves obtained simultaneously during the burst
``forest'' of 2006 March 29. Different energy ranges are shown: 1--4
and 4--10\,keV for the XRT (panels X1 and X2, respectively), and 15--25\,keV,
25--40\,keV, 40--100\,keV and $>$100\,keV for the BAT (panels B1, B2, B3, and
B4,
respectively). The XRT light curves were
background-subtracted and corrected for vignetting, PSF losses,
as well as for pile-up effects. It is evident from the
comparison between XRT and BAT light curves that, on average, the IFs are
harder than the short bursts, though notable exceptions are present.}
\label{fig:lcurve}
\end{center}
\end{figure}

\begin{figure}
\begin{center}
\epsscale{.90}
\includegraphics[angle=-90,scale=.60]{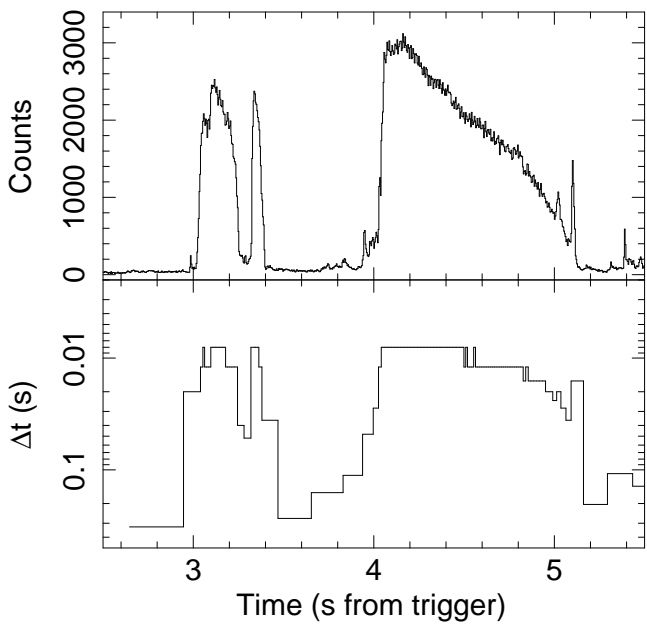}
\caption{Outcoming of the our selection criterion of relying upon a
minimum of 4000 photons in each spectrum. For the first three
seconds of the  2006 March 29 ``forest'' we reported the net counts
(upper panel) and the integration time (lower panel).}
\label{fig:cthresh}
\end{center}
\end{figure}

Burst activity in SGRs and, in particular, in \src\ occurs sporadically 
and is quite diverse in behavior. Burst active phases of SGRs vary both in
total energy released and duration, tending to be concentrated into relatively
narrow intervals (weeks or months) separated by relatively long periods (years)
of quiescence. Since its discovery in 1979, when bursts where detected three
times
in three days, \src\ remained in quiescence until 1992, when a handful of bursts
were again detected within a few days (Mazets \& Golenetskii 1981; Kouveliotou
et
al 1993). In 1998 the source entered an unprecedented level of activity during
which more than 1000 burst were recorded within 9 months, and culminated with
the $\sim 400$\,s long and rather intense 1998 August 27 giant flare event
($\sim4 \times 10^{44}\,\ergs$). Burst activity was detected until 1999.
Particularly intersting, are two events detected in May and
September 1998, during which "bunching" of short and long burst were detected
(Hurley et al. 1999; Mazets et al. 1999). Finally, after almost two years of
inactivity, on 2001 April 18, an intense and long ($\sim40$\,s)
burst was detected from \src\ (Guidorzi et al.\ 2004). The energy released by
this flare (few $\times$ $10^{42}$\,ergs) was less than that of the giant flare,
though larger than that of the most common short SGR bursts. Consequently the
flare was dubbed an ``intermediate flare'' (Kouveliotou et al.\ 2001).
Indeed, looking back at the burst history of \src, it
was realized that a handful of similar events, characterized by a longer
duration (few seconds--few tens of seconds) and a higher fluence than
those characterizing short bursts, were present. These bursts were then grouped
into the class of intermediate flares  likely forming a continuum in
terms of duration and fluences (Olive et al.\ 2004). 

A number of broad-band spectroscopic studies of short and intermediate flares
were carried out in the past by different missions. These found that, above
15--20\,keV spectra are usually well fitted by an optically-thin thermal
bremsstrahlung (OTTB)  with characteristic temperatures ranging from 20 to
40\,keV (Aptekar et al.\ 2001).  Laros et al.\ (1986) and Fenimore et al.\
(1994)
studied the cumulative spectral properties of 100 bursts from \sgra\ in
the 5--200
keV energy range finding that the number of the photons in the spectrum of the
detected bursts was remarkably stable in spite of the large intensity spread (a
factor of 50) and that the low-energy (below 15 keV) data were inconsistent with
the back-extrapolation of an OTTB model that provided a good fit to the
high-energy portion of the spectrum.  Qualitatively similar spectral properties
were measured during a bright intermediate flare from \src\ using {\em HETE--2}
(Olive
et al. 2004) data, which also showed that the OTTB model largely overestimated
the flux
at low energies ($<$15\,keV). The broadband spectrum (7--100\,keV) was best
fitted by the sum of two blackbody (BB) laws. Similar conclusions were reported
by Strohmayer \& Ibrahim (2000) confirming the OTTB limits below 15\,keV  by
using data from the Rossi X-Ray Timing Explorer.

Feroci et al.\ (2004) analyzed the 1.5--100\,keV BeppoSAX spectral properties 
of
10 short burst from \src, further confirming that the widely used optically-thin
thermal-bremsstrahlung model provides acceptable spectral fits for energies
higher than 15\,keV but severely overestimates the flux at lower energies.  A
better fit was obtained by means of an alternative spectral model such as two
blackbodies or a cutoff power law. Most of these studies were hampered by one or
more of the following limitations: relatively poor statistics, narrow
energy bandpass, poor time resolution, off-axis data, and low
sensitivity of wide field of view
instruments.  

Cumulative analysis of 50 bursts detected by {\em HETE-2} from \src\
showed the presence of a time lag of about 2.2$\pm$0.4\,ms between the
30--100\,keV and 2--10\,keV radiation bands implying a rapid spectral softening
and energy re-injection mechanism (Nakagawa et al.\ 2007). Alternatively, the
softer spectral component could be reprocessed radiation from the harder
emission which might be generated near the neutron star surface. More generally,
the observed time lag favors spectral models with (at least) two components.
In any case the temperatures of the 2BB model does not seem to depend on either
the burst instensity or morphology.

It has been assessed that burst activity in the SGRs can have a persistent
effect on the underlying X--ray source (Woods \& Thompson 2006 and references
therein). During the 1998 burst activity phase of \src, the X--ray counterpart
became brighter, its energy spectrum was altered, and the pulse shape changed
dramatically. A similar behavior was also recorded from \sgra\ when resumed
its burst activity phase in 2004 culminating with the 2004 December 27th giant
flare, and later during the 2005-2006 decaying phase (Woods et al. 2007). It
has also been proposed that the bursting activity phases are related to the
magnetosphere. The correlation between the persistent X-ray flux and the
spectral hardness observed in several magnetar candidates on timescales of years
could  be accounted for if the evolution is regulated by varying ``twist'' of
the magnetosphere, as in the case of \rxj\ (Rea et al.\ 2005a
and Israel et al.\ 2007b) and \sgra\  (Rea et al.\ 2005b and Woods et al.\
2007). As discussed by Thompson et al.\ (2002), magnetars may differ from
standard radio pulsars because their magnetic field is globally twisted inside
the star, up to a strength of about 10 times the external dipole and,
occasionally, some of its helicity propagates across the NS surface through
crustquakes, leading to an impulsive twist up of the external field, as well. 
Such an evolving magnetic field is responsible for the different forms
of activity and also for the glitches.
The basic idea of the ``twist'' scenario is that when a static twist is 
formed, currents flow into the magnetosphere. As the twist angle
$\Delta\phi_\mathrm{NS}$ grows, electrons provide an increasing optical depth to
resonant cyclotron scattering, leading to the build-up of a flatter
photon power-law component. At the same time, returning currents produce
an extra heating of the star surface and an increased X-ray flux. The few
long-term X--ray monitoring observations of magnetars collected until now are
consistent with a scenario in which the twist angle steadily increased
before the glitch epochs, culminating with glitches, sometimes bursts, and  a
period of increased timing noise, and then decreased, leading to a smaller
flux and a softer spectrum. 

In this paper, we report the results obtained from our analysis of the data
collected by the \swift\ Burst Alert Telescope (BAT; Barthelmy et al.\ 2005) and
X--Ray Telescope (XRT; Burrows et al.\ 2005) which monitored \src\ during a
burst active phase in 2006 March (Palmer et al.\ 2006). \src\ triggered BAT five
times in a few days at the end of 2006 March, and once a couple of months later.
In particular, on March 29 (Romano et al.\ 2006a) an extremely intense emission
event (fluence of $\sim2$--$3\times10^{42}$\,ergs), lasting $\sim$30\,s, 
of short and relatively energetic bursts was recorded while both BAT and XRT
were simultaneously observing the source thanks to an automatic slew performed
in response to a previous trigger. About 40 single bursts were detected during
this ``forest'', or  ``storm'', a handful are long enough ($\geq$500ms) to be
considered ``intermediate flares'' (IFs). 
The unprecedented BAT and XRT statistics allowed us to perform both average and
time-resolved spectral analysis in the 1--100\,keV band, the first ever on a
large sample of intermediate flares on ms-timescales. We also studied the
persistent emission of \src\ preceding and following the ``forest'' in search
for transient and permanent spectral variations on timescales of days and weeks,
timescales which have been studied rarely. The timing properties of the event
are discussed in an accompanying paper (D.\ Palmer et al.\ 2008, in
preparation).
\begin{table}
\caption{Summary of the average reduced $\chi^2$ and standard deviation 
for different single- and multi-component models obtained for the 729
time-resolved
BAT spectra and the 8 BAT$+$XRT spectra of the most intense bursts.} 
\begin{center}
\label{XRTBATspectra}
\begin{tabular}{lccccc}
\\ \tableline\tableline
Spectral                  &   $<\chi^2_{\nu}>$ &   $\sigma$ &
$<\chi^2_{\nu}>$ &   $\sigma$ & Parameters \\
Model &  \multicolumn{2}{c}{BAT+XRT\tablenotemark{a}} & \multicolumn{2}{c}{BAT}
& (\#) \\
\tableline
Bremss                  &         4.84        &       1.17	  &  1.71    &  
1.95 & 2 \\
DiskBB              &         2.91          &     0.84	  &   1.17   &   
0.51 & 2\\
CompST              & 2.41&0.16	&   1.08   & 0.42 & 3\\
CutoffPL             &  1.36  &  0.07	  & 1.07    &0.23 & 3 
\\
Bremss+BB      	  &  1.33	  & 0.25   &  1.06 & 0.28 & 4 \\
CompTT               &  0.88 &  0.07 & 0.99  &  0.17 & 4 \\
BB+BB               & 0.88 & 0.08&  1.01   & 0.16& 4 \\
\tableline
Bremss\tablenotemark{b}		&    \nodata    & \nodata  &   0.91  &    0.11 &
2 \\
\tableline

\end{tabular}
\end{center}
   \tablenotetext{a}{We used a free scaling factor between XRT and BAT of the
order of 10\%.} 
    \tablenotetext{b}{Fit carried out in the 15--50\,keV range.} 
\end{table}

\section{Observations and Data Reduction}
Table~\ref{sgr:alldata} reports the log of the \swift\ observations of \src\ 
that were used for this work.  The BAT data of the bursts were
accumulated in event mode, with full temporal and spectral resolution,
and archived in sequences labeled by the BAT trigger number (the first six
sequences in Table~\ref{sgr:alldata}). The sequence corresponding to the March
29 ``forest'' is 00203127000.
The XRT data of pointed observations were archived 
in the following 13 sequences of Table~\ref{sgr:alldata},
which span from 2006 March 25 to 2006 June 10.
Note that the XRT data of the ``forest'' are stored in sequence 0030386003.
\begin{figure*}
\begin{center}
\epsscale{.90}
\includegraphics[angle=-90,scale=.55]{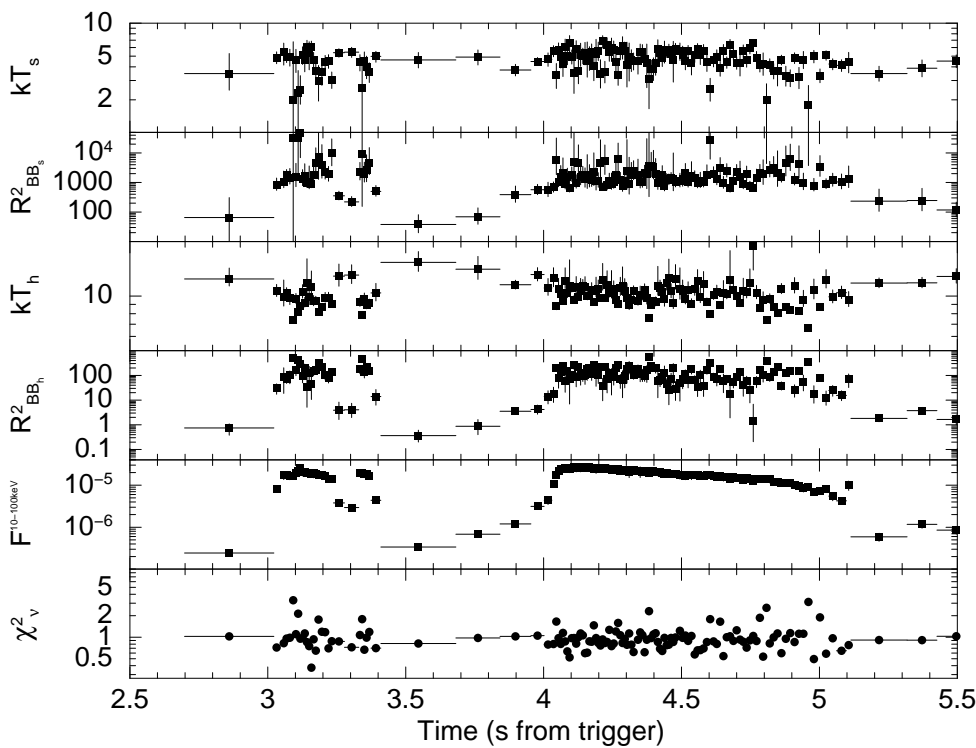}
\caption{Time-resolved results from the spectral fits obtained in the
case of the {\em
BB+BB} model for the first three seconds of the BAT light curve. The soft BB
temperature and radius, the hard BB temperature and radius are reported
together with the observed flux in the 10--100\,keV band and the reduced
$\chi^2$ (from the top to the bottom panel).}
\label{fig:ex_2bb}
\end{center}
\end{figure*}

The BAT data were analyzed using the standard BAT analysis 
software distributed within FTOOLS v6.0.5. The arrival times of events were 
converted to arrival times in the reference frame of the Solar System Barycenter
(SSB) using the {\tt barycorr}  task.  Mask weighted (i.e.\ background
subtracted) light curves were created from
event data in the 15$-$25, 25$-$40, 40$-$100, and 15$-$100~keV energy ranges at
1\,ms and 4\,ms time resolution (see Figure~\ref{fig:BATlcurve} and
the four lower panels of Figure~\ref{fig:lcurve} for the light curves of the
March 29 ``forest'').
Response matrices for BAT spectra were generated with the task  {\tt batdrmgen}
using the latest spectral redistribution matrices.  For our spectral fitting
(XSPECv12.2.1) we considered the 14$-$100\,keV energy  range and applied an
energy-dependent systematic error vector\footnote{
http://heasarc.gsfc.nasa.gov/docs/swift/analysis/bat\_digest.html}.
The time resolution of BAT is 100 microsecs, so pile-up is never an issue
under normal circumstances (see Barthelmy et al. 2005). No deadtime effects 
have been found for these observations (BAT team; private communication). 

\begin{figure*}
\begin{center}
\epsscale{.90}
\includegraphics[angle=-90,scale=.72]{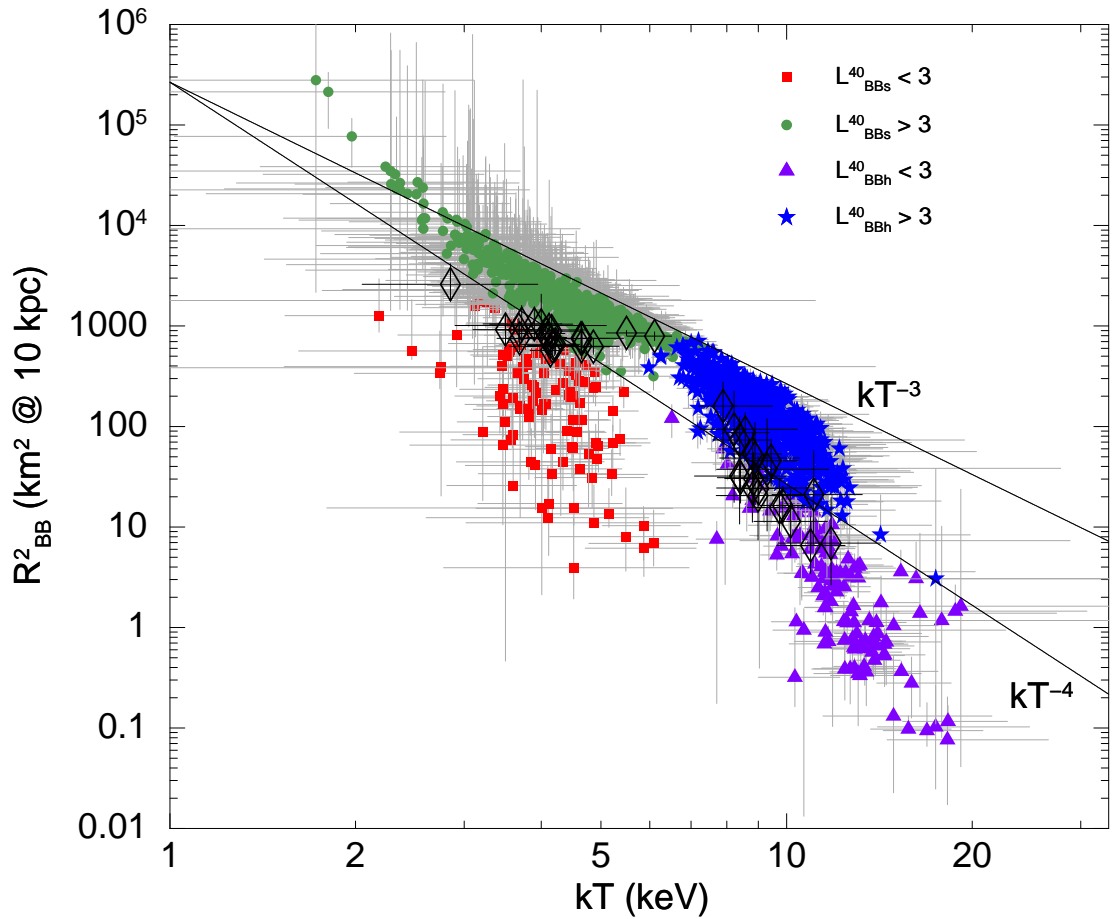}
\caption{Square of radii of the two blackbodies as a function of their
temperatures for the whole $\sim$30\,s BAT dataset. Red filled squares and
dark green filled circles mark the BB$_{\rm s}$ component for luminosities
below and
above 3$\times$10$^{40}\,\ergs$, respectively. Violet filled triangles and blue
filled stars are the same quantities as above for the BB$_{\rm h}$ component.
Black
diamonds marks the Olive et al. (2004) measurements obtained for the 2001 IF. 
We also plot two representative power-laws for $R^2=kT^{\alpha}$, with
$\alpha$ equal $-3$ and $-4$ (the latter corresponding to the expected relation
for a pure BB.)}
\label{fig:2bb_3D}
\end{center}
\end{figure*}

\begin{figure}
\begin{center}
\epsscale{.90}
\includegraphics[angle=-90,scale=.55]{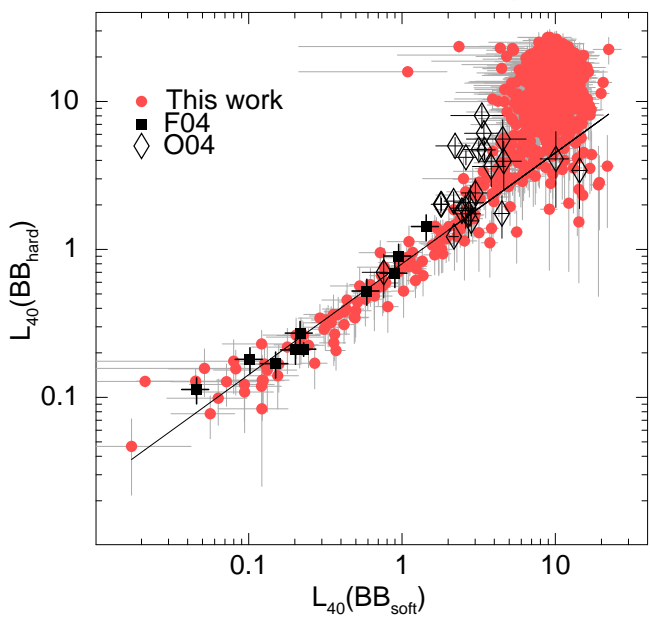}
\caption{Time-resolved bolometric luminosity of the BB$_{\rm soft}$ vs.\ that of
the
BB$_{\rm hard}$ (in units of 10$^{40}$). The solid line marks the
power-law relation $L_{\rm soft}^{40}=(L_{\rm hard}^{40})^{\alpha}$
with $\alpha$=0.70$\pm$0.03  (1$\sigma$ c.l.) obtained by fitting  points
in the 0.01--3 $L_{\rm soft}^{40}$ range. Filled squares represent the
measurements by Feroci et al.\ (2004) for short bursts.}
\label{fig:Ls_vs_Lh}
\end{center}
\end{figure}

The XRT data were first processed by the Swift Data Center at NASA/GSFC into
Level 1 products (calibrated and quality-flagged event lists). Then they were
further processed with the {\tt Heasoft} (v6.0.5)  to produce the final cleaned
event lists. In particular, we ran the task {\tt xrtpipeline} (v0.10.3) applying
standard filtering and screening criteria, i.e., we cut out temporal intervals
during which the CCD temperature was higher than $-47$ $^\circ$C, and we removed
hot and flickering pixels which are present because the CCD operated at a
temperature higher than the design temperature of $-100$ $^\circ$C due to a
failure in the active cooling system.  An on-board event threshold of $\sim$0.2
keV was also applied to the  central pixel, which has been proved to reduce most
of the background  due to either the bright Earth limb or the CCD dark current
(which depends on the CCD temperature).  

For our analysis we considered Windowed Timing (WT) and Photon Counting (PC)
mode data (see Hill et al.\ 2004 for a full description of read-out modes) 
and further selected XRT event grades 0--12 and 0--2 for WT and PC data,
respectively (see Burrows et al.\ 2005). The arrival times of
events were converted to arrival times in the reference frame of the SSB. 

During sequence 00030386003 (i.e. the ``forest'') the count rate of the outburst
was high enough  to cause pile-up even in the WT mode data.  
Therefore, to account for this effect, the WT data were  extracted in a 
rectangular 40$\times$20-pixel region with a 4$\times$20- or 6$\times$20-pixel 
region excluded from its center depending on the peak source intensity, 
i.e.\  with a 4$\times$20-pixel exclusion region during the first and
third snapshot\footnote{A snapshot is a continuous pointing at the target.} 
(where the observed count rate reached $\approx 350$ counts s$^{-1}$), 
and  with a 6$\times$20-pixel exclusion region during the second
snapshot
(observed count rate reaching $\approx 1300$\,counts s$^{-1}$). 
The size of the exclusion region was determined following the 
procedure illustrated in Romano et al.\ (2006b) and correspond to 30--39\% 
(6--4 pixel hole) of the XRT PSF.  
Ancillary response files were generated with the task {\tt xrtmkarf}
within FTOOLS, and account for different extraction regions and  PSF
corrections. We used the latest spectral redistribution matrices in the
Calibration  Database maintained by HEASARC.

The PC data show an average count rate of $\sim 0.08$ counts s$^{-1}$
throughout the entire monitoring campaign, therefore no pile-up correction was
necessary.\ We extracted the source events in a circle with a radius of
20--pixels  ($\sim47$\arcsec).  To account for the background, we extracted WT
events within a  rectangular box (40$\times$20 pixels), and PC events within an
annular region  (radii 85 and 110 pixels) centered on the source and 
far from background sources. 

The energy-resolved light curves during the burst active phase is shown in
Figure \ref{fig:lcurve} both for the XRT (panels marked with X1 and X2) and the
BAT
(panels B1 to B4). The XRT light curves were background-subtracted
and corrected for vignetting and PSF losses, as well as for pile-up.

        \subsection{Time-resolved BAT Spectroscopy}

This section refers to the analysis of the BAT dataset
recorded after the fourth  trigger (sequence 00203127000; 
see Table~\ref{sgr:alldata}).
In consideration of  the extremely pronounced variability of the source during
the burst active phase we adopted the following strategy for the time-resolved
spectroscopic analysis: we selected a 4000-count threshold for the 
accumulation of each spectrum. This resulted in a
set of 729 BAT mask weighted (i.e.\ background subtracted) 
spectra extracted from the event data of sequence 00203127000. The choice of
the threshold is our best tradoff at optimizing the S/N of the spectra, while
still closely following the sharpest features of the light curve. 
Figure~\ref{fig:cthresh} shows the results of our spectral accumulation
criterion for the first three seconds of the BAT light curve. In particular,
the choice of 4000 counts per spectrum is  such that (i) the rapid increase of
the first IF is well sampled, (ii) the first two short bursts (peaking at about
3.10\,s and 3.35\,s) are resolved in time, (iii)  90\% of the IF decay is
sampled with a time resolution of $\leq$10\,ms. All spectral
channels with poor statistics (below 20 counts) were removed before fitting.
The resulting accumulation time of the BAT spectra ranges from 8\,ms to
about 400\,ms.
\begin{deluxetable*}{llllrlcrll}
  \tablewidth{0pc}
  \tabletypesize{\scriptsize} 
  \tablecaption{XRT WT and BAT spectral fit
results of the 7 identified IFs and one short
burst (SB) for the {\em BB+BB} model. Fluxes  are
not corrected for the $N_{\rm H}$. 
\label{IF_spectral_parameters}} 
  \tablehead{	
\colhead{Bin} & \colhead{$N_{\rm H}$} & \colhead{$kT$} & 	
\colhead{$R_{\rm BBs}$} & \colhead{$kTh$} & \colhead{$R_{\rm BBs}$}
& \colhead{$F^{1-10\,{\rm keV}}_{X}$} 
& \colhead{$F^{10-100 \,{\rm keV}}_{X}$} 
& \colhead{$\chi^2_{\rm red}$} &  \colhead{d.o.f.} \\
\colhead{} & 	\colhead{(10$^{22}$ cm$^{-2}$)} & 	\colhead{(keV)} & 
\colhead{(km)\tablenotemark{a}} & \colhead{(keV)} & 
\colhead{(km)\tablenotemark{a}} & \multicolumn{2}{c}{($\times 10^{-7}$ ergs
cm$^{-2}$ s$^{-1}$)}
& \colhead{} & 	  \colhead{}
}
  \startdata	
IF1	&	$2.0\pm0.3$	&	$6.2\pm0.3$	&	$15.9\pm4.2$
&	$12.0\pm0.3$	&	$3.5\pm1.2$	&	$4.68\pm0.15$	&
$103.1\pm1.8$	&	0.84	&	135	\\
IF2	&	$1.9\pm0.3$	&	$5.6\pm0.3$	&	$17.6\pm4.7$
&	$10.4\pm0.3$	&	$3.6\pm1.5$	&	$4.77\pm0.15$	&
$64.4\pm1.7$	&	1.15	&	153	\\
SB	&	$1.2\pm0.3$	&	$5.5\pm0.3$	&	$13.4\pm4.0$
&	$12.1\pm0.4$	&	$2.2\pm0.8$	&	$2.71\pm0.13$	&
$29.1\pm0.9$	&	0.77	&	105	\\
IF3	&	$2.1\pm0.6$	&	$5.4\pm0.3$	&	$17.1\pm5.4$
&	$10.2\pm0.4$	&	$3.6\pm1.6$	&	$4.29\pm0.24$	&
$74.5\pm1.9$	&	0.97	&	100	\\
IF4	&	$2.1\pm0.5$	&	$6.1\pm0.3$	&	$17.4\pm4.9$
&	$11.7\pm0.3$	&	$3.7\pm1.4$	&	$5.49\pm0.23$	&
$127.8\pm3.0$	&	0.88	&	115	\\
IF5	&	$2.8\pm0.4$	&	$6.1\pm0.3$	&	$17.1\pm4.6$
&	$11.0\pm0.3$	&	$4.1\pm1.6$	&	$5.19\pm0.17$	&
$89.8\pm2.4$	&	1.07	&	146	\\
IF6	&	$2.3\pm0.4$	&	$5.1\pm0.3$	&	$19.0\pm5.9$
&	$9.8\pm0.3$	&	$3.8\pm1.6$	&	$4.70\pm0.18$	&
$53.5\pm1.9$	&	0.73	&	117	\\
IF7	&	$2.7\pm0.5$	&	$5.7\pm0.3$	&	$17.1\pm5.1$
&	$11.6\pm0.3$	&	$3.6\pm1.2$	&	$4.71\pm0.20$	&
$85.1\pm2.8$	&	0.77	&	116	
    \enddata 
    \tablenotetext{a}{Assuming a distance to the source of 10\,kpc. } 
   \end{deluxetable*}  

Several different spectral models were adopted in the fit of the whole sample of
729 spectra in an automatic fashion. For each spectrum and model we
recorded the values of the best-fit parameters, unabsorbed flux, and 
reduced $\chi^2$. Then, we averaged the measured $\chi^2$ and inferred their
standard deviations. The latter quantities can be considered as a qualitative 
indication of the spectral model fit goodness, on average. The results are given
in Table~2 ordered with the increasing number of degrees of freedom in the
adopted model. 

Similarly to previous studies reported above and in the literature,
we tested both single and multi-component models. Former models include: an
OTTB ({\em Bremss}) given that it was found to well describe the hard X-ray
spectra of SGR bursts, a power-law with an exponential cut-off ({\em CutoffPL})
which may be considered an extension of the simple OTTB, a disk blackbody ({\em
DiskBB}) which is a generalization of a thermal spectrum including also
geometrical parameters, and different versions of a Comptonized spectrum (such
as {\em CompST}) representative of models involving a thermal spectrum modified
by  mechanisms which preserve the photon number. Among double component models 
we considered a two blackbody ({\em BB+BB}) able to fit well the broad band
energy spectra of both short and long SGR burst, an OTTB with a soft blackbody
in order to adjust the OTTB model extrapolation at low energy, and a
Comptonization model where soft photons are up-scattered in a hot plasma taking
into account relativistic effects ({\em CompTT}; Titarchuk 1994) which is, as a
first approximation, a generalization of both the {\em BB+BB} model (in the case
of saturated Comptonization) and of the {\em CompST} one. 

The procedure of increasing new spectral components (therefore increasing the
number
of free parameters) was stopped when the F-test probability for the inclusion of
an additional component became less than 3$\sigma$. It is evident from
Table~2 that small (close to unity) reduced $\chi^2$ are obtained, for BAT
spectra, for one or two component models with free parameters $\geq$ 3.
Moreover, small reduced $\chi^2$ are obtained, for BAT+XRT joined spectra, only
for two component models with 4 free parameters. The inclusion of a third
component (5 or 6 free parameters) is not statistically significant. 

The {\em CompTT} and  {\em BB+BB} are by far those that 
gave the best reduced $\chi^2$ (average of 0.99 and 1.01). Moreover, as already
reported by
several authors, while the OTTB model fits the spectra well on
relatively narrow energy ranges (reduced $\chi^2\approx0.9$ in the 15--50 keV
range), it fails to produce and acceptable fit once a broader
energy interval is considered (reduced $\chi^2\approx1.7$ in the 
14-100\,keV energy range). A better description is obtained by adding a BB
component to the OTTB spectrum (reduced $\chi^2\approx1.07$). This additional BB
component turned out to have a characteristic temperature  similar to that of
the softer BB component in the 2BB model. 

The average temperatures and sizes of the BB components in the {\em BB+BB} 
during the IFs is $<$kT$_{\rm soft}$$>$$=4.8\pm$0.3\,keV with $<$R$_{\rm
soft}$$>$$=30\pm2$\,km, and $<$kT$_{\rm hard}$$>$$=9.0\pm0.3$\,keV with
$<$R$_{\rm
hard}$$>$$=5.7\pm0.5$\,km, which are in good agreement with previous results
based
on an IF (Olive et al. 2004) and a large  sample  of short bursts (Nakagawa et
al. 2007). However, with respect to the latter cases, the present \swift\ data
represent the best  opportunity (so far)  to study the above reported
spectral parameters as a function of time with good statistics
coupled to fine timing resolution. Figure \ref{fig:ex_2bb} shows, as an
example, the results of the automatic spectral fitting in the case of the 2BB
model for the first 3\,s of the high time resolution BAT data, while in Figures
\ref{fig:2bb_3D} and \ref{fig:Ls_vs_Lh}, we used the inferred values for the
entire 30\,s duration and tried to characterize the main properties of the
adopted model.

        \subsection{Time-resolved BAT+XRT Spectroscopy}
Joint fit of the BAT and simultaneous XRT/WT data of the bursting phase 
associated with the fourth trigger was performed with a second
set of eight BAT and XRT/WT spectra (sequences 00203127000 and 00030386003).
The eight time intervals during which these spectra were accumulated  correspond
the occurrence of  the seven IFs plus a canonical short burst (as a comparison).
We report the main spectral parameters in Table~2 and
Table~\ref{IF_spectral_parameters}.
 
The results of the XRT plus BAT fits (see Table~2) confirm those obtained with 
the BAT in the previous section: the  {\em CompTT} and the {\em BB+BB} models
gave the best reduced $\chi^2$ ($\approx0.9$), the {\em OTTB} ({\em Bremss} 
Table~2 and hereafter) the worst. However, when a
BB component is added to the latter component a better solution is obtained
(with a number of d.o.f. equal to that of the {\em CompTT} and the {\em BB+BB}
models, though the reduced $\chi^2$ ($\approx$1.3; Table~2) remains too high to
be acceptable. More generally, the addition of a BB component
to the other poorly fitted models, such as the {\em CompST} and {\em 
CutoffPL}, results in a smaller reduced $\chi^2$, though not yet acceptable
and with a larger number of free parameters (5). In all cases the
characteristic temperature of the soft thermal component is similar to that of
the soft BB in the 2BB model. These findings, together with similar results
obtained by the analysis of broad band spectra of SGR bursts from other
missions, in slightly different (but overlapping) energy intervals,
and by using different procedures/algorithms, make us confident that the \swift\
fitting results are reliable.


We note that adopting a different abundance than solar values for the absorption
multiplicative component  (Anders \& Grevesse 1989; Lodders 2003) does not
significantly modify the spectral parameters for the joint XRT plus BAT spectral
fit. 
\begin{figure}
\begin{center}
\epsscale{.90}
\includegraphics[angle=-90,scale=.40]{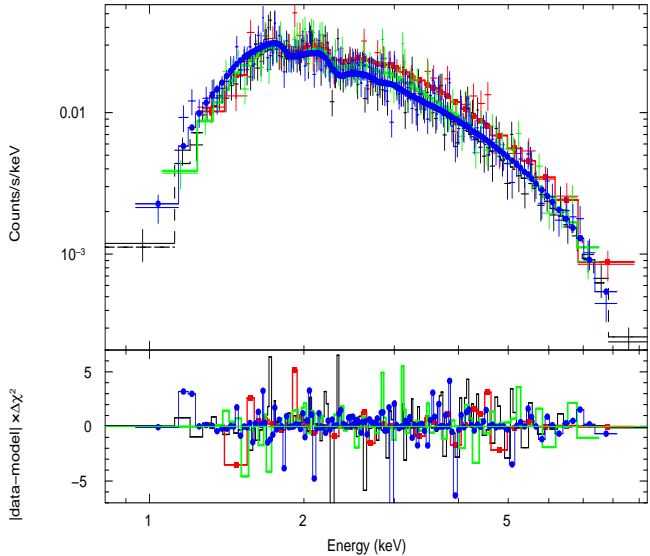}
\caption{The 1--10 keV Swift XRT spectra (PC mode only) of the persistent
emission of \src\ are shown together with the model (BB plus PL) residuals. 
The corresponding \swift\ data sequences are 00030386001 (black lines),
00030386002  (red lines and filled squares), 00030386009 (green lines),
and 00030386010--6014 (blue lines and filled circles).}
\label{fig:spectrum_quiesc}
\end{center}
\end{figure}

\begin{figure}
\begin{center}
\epsscale{.90}
\includegraphics[angle=-90,scale=.42]{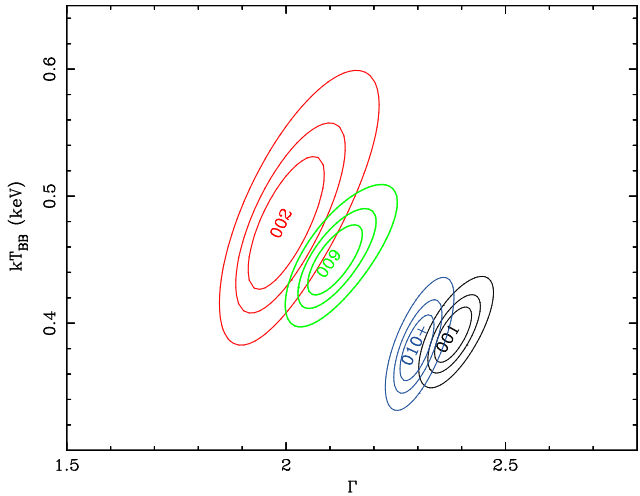}
\caption{The 1, 2, and 3$\sigma$ confidence regions obtained for the BB
temperature $kT$ and the power-law photon index $\Gamma$ for
each spectrum during persistent emission from \src. Note that
the temperature $kT$ is consistent with being constant, while $\Gamma$ is
variable, and harder than in quiescence, at more than 3$\sigma$ level. }
\label{fig:spectrum_contplot}
\end{center}
\end{figure}

\begin{figure}
\begin{center}
\epsscale{.90}
\includegraphics[angle=-90,scale=.42]{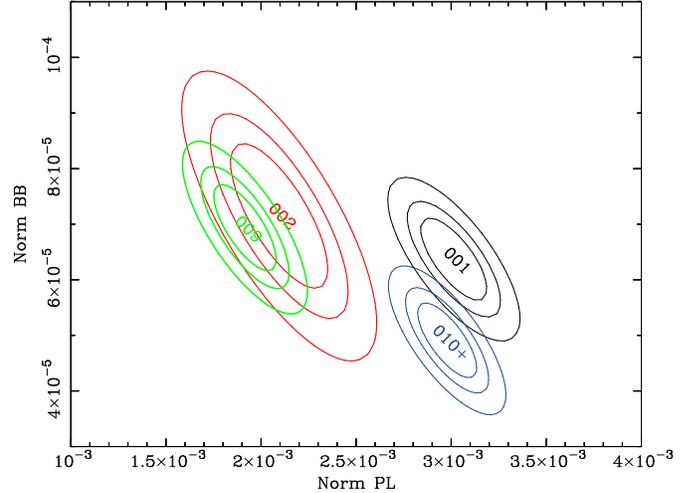}
\caption{The 1, 2, and 3$\sigma$ confidence regions obtained for the BB and PL
normalizations for each spectrum during persistent emission from \src. Similarly
to Figure \ref{fig:spectrum_contplot} the BB component appears to be
approximately constant, while the PL normalization is variable and higher than
in quiescence, at
more than 3$\sigma$ level. }
\label{fig:spectrum_contplot2}
\end{center}
\end{figure}

	\subsection{XRT Monitoring of the Persistent Emission}
        \label{section:xrttimeresspec} 

The approximately two-month time span covered by the \swift\ monitoring
observations was also used to study the timing and spectral properties of the
persistent emission. Only the PC data were used for spectral analysis in order
to rely
upon a better signal to noise ratio. WT data were also accumulated and used for
timing analysis, after removing bursts, in order to rely upon a higher time
resolution. The corresponding \swift\ data sequences used to carry out the
spectral analysis are 00030386001 ($\approx 50$\,ks effective exposure time),
00030386002 ($\approx 11$\,ks), 00030386009 ($\approx 29$\,ks), and
00030386010--6013 ($\approx 51$\,ks). In order to minimize the possible burst
contamination in PC mode, spectra were accumulated after having
excluded $\sim$10s time-interval just before and after each good time intervals
(GTIs) corresponding to automatic switches between PC and WT modes. In fact, 
these switches are likely due to high count rate events from cosmic rays and/or
from genuine bursts from \src. A comparison between the filtered and unfiltered
spectra did not result in the detection of significant differences, making us
confident that burst contamination, if any, is negligible. As in the previous
cases, the data were rebinned with a minimum of 20 counts per energy bin to
allow $\chi^2$ fitting within {\tt XSPEC} (v12.2.1). We further removed those
energy channels with poor statistics after background subtraction.

We assumed an absorbed power-law plus blackbody model as inferred with 
deeper \XMM\ observations (Mereghetti et al. 2006). 
We left the absorption free to vary but forced to be the same in  
different spectra and obtained $N_{\rm H}=(2.3 \pm 0.1) \times 10^{22}$
cm$^{-2}$. The results of the fits are reported in
Table\,\ref{P_spectral_parameters}, while fitted spectra are shown in
Figure\,\ref{fig:spectrum_quiesc} together with their residuals expressed in
units of $\sigma$. There is marginal evidence of parameter variability 
as a function of observed flux: the spectrum becomes harder for larger flux
levels. In order to properly quantify such a spectral variability and its flux
dependence, we carried out a bi-dimensional fit (contour plot) of the main
spectral parameters which has also the advantage of better taking into account
the possible correlations among them. The results of these additional fits are
shown in Figure\,\ref{fig:spectrum_contplot} and \ref{fig:spectrum_contplot2}. 
The ellipsoidal shape of the 1, 2,  and 3$\sigma$ confidence regions (the
innermost region corresponding to 1$\sigma$) clearly indicates that parameters
(and, more generally, the BB and PL components) correlate with each other.
Additionally, the two-parameter fits confirm the  spectral evolution (c.l.
$>3\sigma$) of the persistent spectrum of \src\ as a function of flux: 
the source displayed harder spectra just before (March 28) and after
(April 8) the burst ``forest'' event (March 29), and returned back to the
pre-event level in mid-April. These results are similar to those observed in
other magnetar candidates, such as \sgra\ among SGRs and \rxj\ among AXPs, the
main difference being in the timescales on which these spectral variations
occurred: years for the latter two sources and days/weeks for \src. In the
previous cases this trend has been modelled in terms of ``twisting'' of the
magnetosphere (Thompson et al. 2002). 

\begin{figure}
\begin{center}
\epsscale{.90}
\includegraphics[angle=-90,scale=.63]{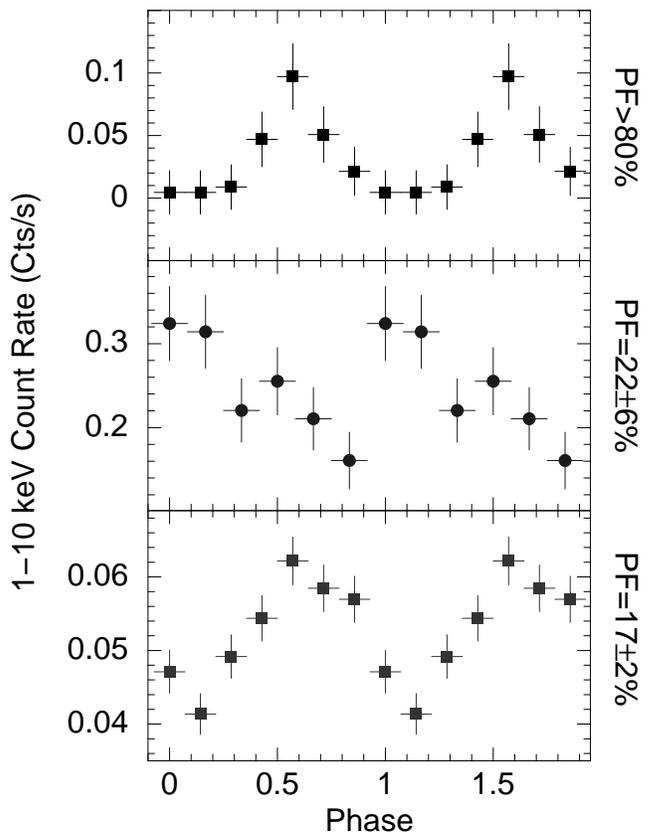}
\caption{The 1--10\,keV \swift\ XRT  lightcurves of \src\ from 
approximately one hour before and after the BAT event (2006 March 29; top and
mid panels), and the first week of April 2006 (bottom panel). In all cases we
used $P$=5.19987$\pm$0.00007\,s, as determined from an \XMM\
observation carried out on 2006 April 1 (Mereghetti et al., 2006). For the
about one week-long baseline of the April 2006 lightcurve dataset we used
$\dot{P}$=(9.2$\pm$0.4)$\times 10^{-11}\,s\,s^{-1}$. The pulsed fraction,
semi-amplitude of modulation divided
by the mean source count rate, is reported on the right of each panel.}
\label{fig:fold}
\end{center}
\end{figure}

The WT and PC XRT data  were also  used to look for the $\sim$5.2\,s
pulsations of \src. Given the limited XRT statistics in the
persistent component and the relatively small pulsed fraction of the pulses
($\sim$15--20\% level; semi-amplitude of modulation divided by the mean source
count rate) we used the period value, $P$=5.19987$\pm$0.00007\,s, inferred from
an \XMM\ observation carried out on 2006 April 1 (Mereghetti et al.\ 2006). We
detected the \src\ pulsations on 2006 March 29, the same day as burst ``forest''
event, in two time intervals preceding (March 29 from 01:02 to 01:26 UT)
and following (from 04:22 to 04:42 UT)  the ``forest''  by approximately 1\,hr
(in both cases the signal was detected at about 3$\sigma$ c.l.). By merging all
the XRT observations carried out in March/April (from 30 March 01:07 to 7 April
23:20) and by using $\dot{P}$=(9.2$\pm$0.4)$\times 10^{-11}$\,s\,s$^{-1}$
reported by Mereghetti et al.\ (2006), the pulsations could be detected at a
high significance level. Figure \ref{fig:fold} summarizes the results of the
timing analysis. Both the pulse fractions and shapes inferred from the first 
dataset are different from those obtained from the second and third
datasets, and, more generally, from what observed on average since the discovery
of this source. At the same time,  the pulse shapes of the second and third
datasets are significantly different. The value and shape of the the latest
dataset is virtually equal (to within the uncertainties) to those obtained by
Mereghetti et al.\ (2006) about 3 days after the BAT event.
\begin{deluxetable*}{llllcll}
  \tablewidth{0pc} 	      	
  \tabletypesize{\scriptsize} 
  \tablecaption{XRT PC spectral fit
results of the persistent emission of \src\ during the four time
intervals reported in the text and by assuming the {\em BB+PL} model.
Fluxes are not corrected for the $N_{\rm H}$ value (which is left to vary but
constant among spectra). \label{P_spectral_parameters}} 
  \tablehead{	
\colhead{Interval}  & \colhead{$kT$} & 	
\colhead{$R_{\rm BBs}$} & \colhead{$\Gamma$} &
\colhead{$F^{1-10\,{\rm keV}}_{X}$} 
& \colhead{$\chi^2_{\rm red}$}
& \colhead{d.o.f.} \\
\colhead{MJD} & 	\colhead{(keV)} & 
\colhead{(km)\tablenotemark{a}} & \colhead{} & \colhead{($\times
10^{-12}$ ergs cm$^{-2}$ s$^{-1}$)}
& \colhead{} & 	  \colhead{}
}
  \startdata
53819.9519--53821.9082& 0.39 $\pm$ 0.02  & 3.4 $\pm$  0.5 &  2.4 $\pm$  0.1 &
 4.6 $\pm$
0.8 & 1.13 & 125   \\
53822.0455--53822.9923& 0.51 $\pm$  0.06  &  2.9 $\pm$  0.6  & 1.9 $\pm$  0.3 &
 6.3 $\pm$
1.7&  0.91 &  39   \\
53833.0355--53835.9854& 0.48 $\pm$  0.03  &  3.3 $\pm$  0.4 &  2.1 $\pm$  0.2 
&   5.0 $\pm$
1.4 & 0.96 & 86   \\
53836.0335--53840.0& 0.38 $\pm$  0.02  &  4.3 $\pm$  0.8 &  2.3 $\pm$  0.1 &
 5.0$\pm$ 
0.7&  0.94  &  137  \\
  \enddata 
    \tablenotetext{a}{Assuming a distance to the source of 10kpc and an 
absorption hydrogen column of 2.3$\pm$0.1 $\times$ 10$^{22}$ cm$^{-2}$. } 
   \end{deluxetable*}  

        \subsection{UVOT Observations and Data Reduction}

The Swift UV/Optical Telescope (Roming et al.\ 2005) is a 30\,cm 
modified Ritchey-Chretien telescope co-aligned with the XRT. It features a 
photon-counting micro-channel plate intensified CCD detector optimized for 
UV and optical light over the 170$-$600~nm range; its six broadband
colour filters sample this range.

The UVOT was taking data in the $V$ filter (central wavelength 5460~\AA) 
during the March 29 outbursts of \src. No optical counterpart 
is detected in the resulting image, accumulated between 
2006-03-29 02:39:57 and 2006-03-29 03:06:00 UT with a total exposure time 
of 1539\,s. In the following we derived an upper limit to the optical emission
during the burst ``forest''. Counts were extracted from a 2\arcsec\, radius
aperture at the position of the source in the XRT; the size of the aperture was
chosen so as to minimize contamination by a $V \sim$ 18.6 mag star (USNO number 
0975-14353360) which is only 3\arcsec\, from the XRT position. The counts were 
aperture-corrected up to 6\arcsec, and used to calculate a 3$\sigma$ 
$V$-band magnitude upper limit for any counterpart of 20.7\,mag. Adopting the 
Akerlof et al.\ (2000) estimate for $A_{\rm V}$ of $\sim$ 10 mag for 
the Galactic reddening in the direction of \src\, we estimate a 
dereddened upper limit of 10.7\,mag.
\begin{table}
\begin{center}
\label{UVOT_limits}
\caption{The times of \src\ outbursts occurring when UVOT was taking data,
with the filter in use at the time (central wavelength in brackets), exposure 
time of the image in seconds, 3$\sigma$ magnitude upper limit at the XRT 
position.} 
\begin{tabular}{lrrr}
\\ \tableline\tableline
Flare time              &  Filter            &   Exp   &   UL   \\
\tableline
2006-03-25 23:34:07.748 &  $UVM2$ (2200~\AA) &   642   &   21.3  \\
2006-03-26 00:59:22.680 &  $V$    (5460~\AA) &   213   &   18.9  \\
2006-03-26 13:30:04.620 &  $B$    (4350~\AA) &   216   &   20.8  \\
2006-03-26 16:50:17.632 &  $UVW2$ (1930~\AA) &   851   &   20.9  \\
2006-03-28 20:32:31.456 &  $UVM2$ (2200~\AA) &   261   &   20.8  \\
2006-03-29 02:53:09.46  & $V$    (5460~\AA) & 1539 & 20.7  \\
\tableline
\end{tabular}
\end{center}
\end{table}

For each UVOT image coincident with an outburst, counts were extracted from
4\arcsec\, radius  apertures at the position of the source in the XRT. The
counts were  aperture-corrected up to 6\arcsec\, (optical filters) or
12\arcsec\, (UV filters; the different apertures are due to the broader PSF in
the UV),  and used to calculate 3$\sigma$ magnitude upper limits for any 
counterpart. Using the Akerlof et al. (2000) estimate of $A_{\rm V}$ $\sim$ 
10 magnitudes for the Galactic reddening in the direction of \src, we 
corrected the upper limits for Galactic extinction assuming E(B-V)=3.125. 
The resulting values are listed in Table~5.

\section{Discussion}
\label{section:dis}

\swift\ recorded 
a rather intense and rare series of short bursts and intermediate
flares (the densest part of which lasted for $\sim$30\,s) from \src\ during 2006
March exploiting its rapid follow-up pointing capability. This
allowed us to carry out, for the first time, a detailed study of
the average and time-resolved spectral properties of this source during such
events. Moreover, the continuous monitoring of \src\ on longer timescales,
afforded a study of the changes of the spectral and timing properties of the
persistent emission.  

\subsection{Burst emission}

We used several different spectral models comprising one or
two components, to fit both the time-resolved BAT spectra and the time-averaged
(over a whole single burst or IF) BAT+XRT spectra. Interestingly, only two
({\em CompTT} and 2BBs) out of the five models considered ({\em CompST}, {\em
CutoffPL}, {\em Bremss+BB}, {\em CompTT} and 2BBs) which fit the BAT spectra
well, are also able to model the BAT+XRT spectra (see Table~2
and Table~\ref{IF_spectral_parameters}). An optically-thin thermal
bremsstrahlung ({\em Bremss}) does not yield an acceptable reduced $\chi^2$ in
either cases (BAT and BAT+XRT spectra). This result is consistent with previous 
spectral analyses carried out on broad band spectra of a few short bursts and a
4\,s-long IF (see, Feroci et al. 2004 and Olive et al. 2004). Notably, the
{\em Bremss} model describes the BAT spectra well only when a reduced energy
interval,
15--50\,keV, is considered. In the following we discuss in greater detail the
results inferred from the 2BB model (note that the {\em CompTT} model shape is
virtually identical to that of the 2BB in the limit of saturated
Comptonization). Moreover, in consideration of the difficulties in explaining 
short bursts and IFs in the accretion scenario
(they have extreme super-Eddington luminosities are in fact observed), we
discuss here the
implications of our results within the magnetar scenario. 

With respect to the former studies of the broad-band spectrum 
of similar events, the \swift\ observation of 2006 March 29 has the
advantage of a number of events ($>$40 between short
bursts and IFs), therefore largely improving the statistics over which
the detailed analysis can be carried out. The impact of the increased statistics
and time resolution is evident when comparing the \swift\ results with
previous ones.
On 2001 July 2, the {\em HETE-2} instruments recorded a 3.5\,s IF from
\src\ and a time resolved ($\Delta t>$30\,ms) spectral analysis was carried out
in the 2--100\,keV range (Olive et al. 2004). They found that the 2BB model 
fitted the spectra well, the higher temperature BB evolving in a manner
consistent with a shrinking emitting region, the lower temperature BB showing a
constant radius. They also suggested that, within the magnetar model, the 2BB
model might be an approximation of a more complex multi-temperature spectrum.
Their Figures 5 and 7 can be easily compared with our Figures \ref{fig:ex_2bb}
and \ref{fig:2bb_3D}: the \swift\ data fill in those regions on the $kT$-$R^2$
plane, which remained unexplored with {\em HETE-2}.

Several new properties can be immediately inferred: a)  the \swift\ data
populate, almost homogeneously, all temperatures between $\sim$2 and
12\,keV, and the {\em HETE-2} measurements (black diamonds in
Figure~\ref{fig:2bb_3D}) can be regarded as a subset of them;  b) the
distribution is such that a sharp edge between the populated regions and
the rest of the $kT$-$R^2$ plane is clearly present, with a cut-off (or
sharp turn) in the distribution of $R_{\rm BBh}$ at
$kT_{\rm h}\approx12$--13\,keV. This sharp edge provides an estimate of the
typical size of the relevant emitting regions (by
assuming a reference distance of 10\,kpc): 30--200\,km range for the
BB$_{\rm s}$ component and
3--30\,km range for the BB$_h$ one. c) there is an additional turn in the
BB$_{\rm h}$ component size between 6 and 16\,km, at approximately 10\,keV (see
Figure~\ref{fig:2bb_3D}), and 
d) there is a strong correlation between $kT$ and $R^2_{BB}$ corresponding to
the brightest phases of the IFs, which is consistent with a power-law with index
of about $-3$ (while, were the luminosity the same for all events, 
the relationship expected for a BB component would be y=x$^{-4}$, where $x=kT$ 
and $y=R^2_{BB}$). 

The sharp edge in the distribution of the 2BB parameters suggests that a
saturation effect occurs. In order to investigate this effect
further, we divided the data points of the two BB components in two
samples, below and above a bolometric luminosity of
$L=3\times$10$^{40}\,\ergs$ (see Figure~\ref{fig:2bb_3D}).  The luminosity
threshold was chosen so as to roughly separate the peaks of the bursts
and flares from the inter-burst/flare time intervals. We note that the
{\em HETE-2} data lie just above and below the flux threshold, but never
reach the values corresponding to the peaks and tops of the \swift\ IFs.  
As we can see from Figure~\ref{fig:2bb_3D}, as long as each BB component
is below $\sim3\times10^{40}\,\ergs$, their two luminosities are strongly
correlated ($L(BB_{hard})$ vs $L(BB_{soft})$ is well fitted by a power law
with index 0.7$\pm0.3$ at 1$\sigma$ c.l.). This correlation, which seems
indicative of a physical link between the two BB components, was 
already noticed by Feroci et al.\ (2004) who analyzed the time-averaged
spectral properties of 10 short bursts from \src\ recorded by {\em BeppoSAX} in
April 2001 (filled squares in Figure \ref{fig:Ls_vs_Lh}). It is apparent that
the
{\em BeppoSAX} measurements are fully consistent with those obtained with
\swift\ and can be regarded as a subsample of the latter. On the other hand,
this trend was not reported by Olive et al.\ (2004) for the IF observed by {\em
HETE-2} in 2001 (empty diamonds in Figure \ref{fig:Ls_vs_Lh}). Intriguingly,
the correlation no longer holds above $\sim3\times10^{40}\,\ergs$
for the \swift\ data. Above this value, the luminosity of the BB$_{\rm s}$
component reaches a maximum in the (7--14)$\times$10$^{40}\,\ergs$ range,
while that of the BB$_{\rm h}$ continues to increase up to
$\sim3\times10^{41}\,\ergs$. Clearly, this previously unknown  
behavior strengthens the idea of a saturation mechanism in the
burst emission.

The emission properties of the higher luminosity flares ($L>3 
\times 10^{40}\,\ergs$), combined with the spectral information, places some 
basic constraints on their production mechanism. 
The possible location of the emitting region and the 
main processes likely involved in shaping the spectra have been discussed by
Thompson \& Duncan (1995, hereon TD95).  Lyubarsky (2002) 
studied the propagation of a trapped fireball through the 
ultramagnetized magnetosphere, and computed an approximate emission spectrum
for it. However, the comparison of these predictions 
with the results from our observations or those by Olive et al. (2004) is not 
straightforward. As already noted by Olive et al. (2004), the ``modified
blackbody'' proposed by Lyubarsky (2002) predicts a significant depression below
the blackbody of the higher energy tail, which is not found in the data (the fit
remains good up to energies $\sim$ 100\,keV). Clearly, more detailed spectral
models for the burst are required.

In the following, we will concentrate on new important information
(mainly concerning the brightest part of IFs) which has been inferred 
from
our spectral analysis.  In particular, we discuss in detail three main
findings:  i) the existence of two blackbody components, the harder one
($kT_{\mbox{\small{h}}}\sim 7$--$11$\,keV) having a systematically
smaller radius ($R_{\rm BBh} \leq25$\,km) than the softer one ($R_{\rm
BBs}\sim 25$--$100$\,km);
ii) the existence of a clear correlation between the luminosities of the two
components up to $\sim 3 \times 10^{40}\,\ergs$, above which value the 
luminosity of softer blackbody shows signs of  saturation,  
while that of the harder blackbody still grows up to a few times
$10^{41}\,\ergs$; 
iii) the existence of a correlation between temperature and blackbody 
radius, which holds for the most luminous parts of the flares 
(approximately for $L_{\rm tot}> 4 \times 10^{40}\,\ergs$ ).

The first finding shows that different thermal components originate
from different regions around the NS, ranging from its surface (at $R <
R_{\rm NS}$) to well up in the magnetosphere, at a height of several
stellar radii. The maximum observed flare luminosity is $\sim$3$\times
10^{41}\,\ergs$, attained by the hard blackbody component at an effective
temperature of $\sim$10\,keV and radius of $\sim$15\,km. Interestingly, 
this matches well the magnetic Eddington luminosity,  $L_{{\rm Edd,B}}$, 
at that same radius, for a surface dipole field $\simeq$8$\times
$10$^{14}$\,G (this value of the B-field is very close to that
deduced from the spindown rate of SGR 1900+14), where (cfr. Paczynsky 1992,
TD95):
\begin{equation}
\label{eddmagn}
L_{{\rm Edd,B}}(r)\simeq 2 L_{\mbox{\tiny{Edd}}} \left(\frac{B}{10^{12}
\mbox{\small{G}}}\right)^{4/3}  
\approx 2 \times 10^{40} \left[\frac{B}{B_{
\mbox{\tiny{QED}}}}\right]^{4/3} 
\left(\frac{R}{R_{\mbox{\tiny{NS}}}}\right)^{2/3}
~\mbox{erg s}^{-1}.
\end{equation}
If radiation originated from a trapped hot fireball this would be  in line
with the discussion of Thompson \& Duncan (1995, 2001), according to which 
the radiative efficiency of a magnetic confined fireball 
never exceeds (to within a small factor) the magnetic Eddington flux.  
In fact, ablation of matter from the NS surface would increase the
scattering depth in a super-Eddington radiation field,  thus providing a 
self-regulating mechanism. The radius and temperature that we inferred at  
maximum burst luminosity are also in good agreement with the prediction for the 
emission coming from a trapped fireball (TD95, cfr. their
discussion for the best fit temperature and radius in the case of SGR
1806$-$20).

Most remarkably, we find  that the luminosity of the soft 
component can be larger than 10$^{40}\,\ergs$ out to
R$\sim$100\,km. For R$>$30-40\,km this value exceeds
the magnetic Eddington luminosity at the corresponding radius (for a dipole
field geometry) and suggests that magnetospheric 
confinement plays an active role out to those distances, with magnetic stresses
balancing radiation forces. If this were the case, the
saturation of the soft component at $\simeq 10^{41}\,\ergs$  would be 
related to the magnetic field strength at the relevant radii and its maximum
ability to retain the trapped fireball matter which is subject to very high 
radiation pressure. A detailed investigation of the possible equilibrium 
configurations is beyond the scope of this paper and will be addressed in a 
future work. 

Our second finding [point ii)] implies that, in the luminosity range
$10^{40}$--$10^{41}\,\ergs$, the total radiation energy is divided
almost equally between the two components.
Combining the information of Figures~\ref{fig:2bb_3D} and 
\ref{fig:Ls_vs_Lh} we can see that, for a given luminosity $\leq 
10^{41}\,\ergs$, there are two separate emission regions: 
a smaller and hotter one, whose radiating area suggests emission from (part
of) the NS surface with a relatively high effective temperature (10--12\,keV), 
plus a second, possibly magnetospheric region, with significantly larger
emitting area and lower effective temperature (3--7\,keV). 
The hotter component attains the highest luminosities, 
{\textit{i.e.} $> 10^{41}\,\ergs$}, as its radius grows slightly and the 
effective temperature decreases. On the other hand, the colder component is 
characterized by a minimum blackbody radius of $\sim25$\,km and 
maximum temperature of $\sim 7$\,keV. 

A possible interpretation involves the different way in which 
photons with ordinary (O) or extraordinary (E) polarization mode (a
properties introduced by the presence of a birefringent medium,
such as the magnetic field; M\'esz\'aros et al. 1980) 
propagate across the magnetosphere. Since the scattering cross section of 
E-mode photons is much reduced in presence of strong magnetic fields, E-mode  
photons have a scattering photosphere which is located much closer to 
the NS than that of O-mode photons. On the other hand, in  
supercritical magnetic field ($B> B_{\mbox{\tiny{QED}}}$), 
E-mode photons have a non-negligible probability of splitting (and switching to
the O-mode). The probability is a strong function of energy, so that E-mode
photons of high energy cannot travel far from their scattering photosphere
before splitting into O-mode photons. On the other hand, O-mode photons will be 
entrained with streaming electrons and baryons and advected out to where their 
scattering optical depth becomes $\tau(O) \sim 1$ (TD95; Lyubarsky 2002). 
As a consequence of Compton scattering (and, with lower efficiency, photon
merging), a fraction of O-mode photons can also switch back to the E-mode.
Therefore, the two modes are effectively coupled and are advected at comparable
rates as long as mode switching is efficient: an approximately equal
distribution of energy in the two modes is thus to be expected (TD95; Lyubarsky
2002). This may suggest that the two observed spectral components reflect the
population of photons in the two polarization modes and, thus, the regions from
which they are emitted.  
\begin{figure*}
\centering \includegraphics[angle=-90,scale=.80]{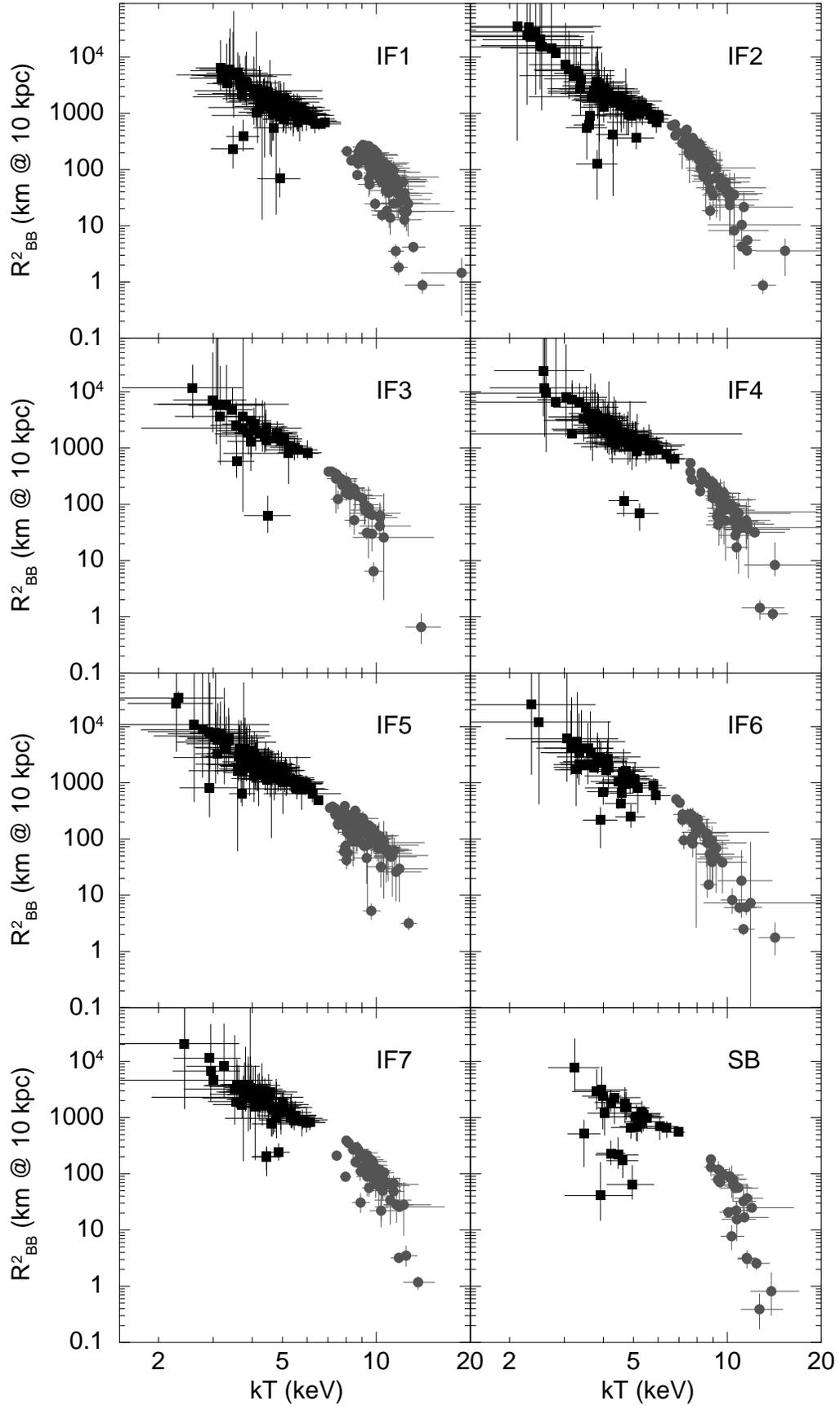}
\caption{Same as Figure\,\ref{fig:2bb_3D} but for each of the seven longest
intermediate flares (IF1 to IF7) plus a short burst (SB) also reported in
Table\,\ref{IF_spectral_parameters}.}
\label{fig:2BB_3D_each}
\end{figure*}

We note that the minimum radius of the cold blackbody almost corresponds to
the maximum radius of the hot component, the distribution of points in the
$R^2$ vs.\ $kT$ plane being continuous across this point (see
Figure\,\ref{fig:2bb_3D}). This
suggests the existence of a narrow zone of separation between the two
emission regions. In order to further characterize this zone, we
studied the $R^2$ vs.\ $kT$ distribution for each of the seven longest IFs
and for one short burst we already discussed in previous sections (see also 
Table\,\ref{IF_spectral_parameters} and Figure\,\ref{fig:2BB_3D_each}).
Remarkably, the transition between the two BB components is more
evident for the brightest flares (IF1 and IF7 in Figure\,\ref{fig:2BB_3D_each})
where the upper left edge of the parameter distribution of the hard component is
confined to smaller radii (R\,$\leq$\,20\,km) and correspondingly higher
temperatures (kT\,$\geq$\,8\,keV). The soft component seems less
dependent on the flux than the hard one, its minimum size being 
of the order of $\geq$\,25-35\,km in all IFs. Qualitatively, this behavior is in
agreement with the above scenario. In fact, a larger brightness can be reached
when the burst energy is trapped closer to the NS surface,
where the stronger magnetic field allows a larger flux (of E-mode photons) to be
released. Moreover, the sharp and narrow gap between the emitting regions of the
two BBs, at a radius of ($\sim20$--30\,km), can be interpreted as the signature
of the presence of an E-mode splitting photosphere (TD95); 
this represents the surface (at radius $R = R_{\mbox{\tiny{QED}}}$) below which
the magnetic field is supercritical ($B \geq B_{\mbox{\tiny{QED}}}$), photon
splitting is efficient and the two polarization modes can remain coupled. E-mode
photons can stream freely to the observer from the splitting 
photosphere (at radius $R_{\mbox{\tiny{QED}}}\sim20$--30\,km), which then would
naturally define a maximum size for the emission region of radiation 
in this polarization mode. On the other hand, O-mode photons could not 
originate from regions internal to the splitting photosphere, as their
scattering optical depth is still much larger than unity within that volume.

If this scenario is correct, it implies that the emission region of E-mode 
photons can range from the base of the trapped fireball (near the
NS surface) to a slightly higher region in the magnetosphere, but well 
within the splitting photosphere. In this case, part of the 
internal energy of the coupled E and O-mode photons is lost in the
E-mode photosphere (adiabatic) expansion from near the NS to the splitting
photosphere. This degrades the original spectrum towards lower temperatures and
larger radii, up to $R_{\mbox{\tiny{QED}}}$. On the other hand, O-mode photons
might either be be released near the splitting photosphere or advected to much
larger radii, depending on the value of their  optical depth at that 
location (see TD95). In any case an improved treatment of the
above effects would be required in order to develop  a self-consistent model.

Finally, our third finding is the apparent correlation between the 
surface and temperature of the spectral components at different luminosities.
This behavior does not have a clear interpretation. As shown in Figure
\ref{fig:2bb_3D}, the top part of the data strip is well fit by a power-law
$\Sigma \propto T^{-3}_{\mbox{\tiny{BB}}}$, where $\Sigma$ is the radiating
blackbody surface. Since $\Sigma T^4_{\mbox{\tiny{BB}}} =L$, this implies
$T_{\mbox{\tiny{BB}}} \propto L$. Therefore, along lines of constant luminosity,
the radiating surface would scale like $T^{-4}$, i.e.  steeper than the data
points (on the edge of the allowed region in the kT vs R$^2$ plane).
The $\Sigma T^3 = $const relation that implies a roughly constant number
of emitted photons per unit time. 

At the highest luminosities ($\geq 10^{41}\,\ergs$) of the flares, where the 
softer blackbody saturates, the observed correlation involves 
both components. It is tempting to attribute this to the fact that, in 
this luminosity range, Comptonization, and/or adiabatic losses play a dominant
role in the formation of the emerging spectrum, both processes being
characterized by conservation of the photon number (photon splitting 
should have only a minor impact). This is in agreement with the prediction by 
TD95, that at effective temperatures less than $\sim10$\,keV photon splitting 
would not be efficient in maintaining a pure Planck spectrum for both modes.
However, if the photon chemical potential is sufficiently small compared
to the photon temperature, deviations from a Planck spectrum are expected only
at low energies (TD95; Lyubarski 2002), which we do not observe given they are
masked by absorption.

On the other hand, at luminosities $\leq 10^{41}\,\ergs$, the two 
blackbodies approximately share the same luminosity, therefore the colder one
is emits a larger number of photons. This suggests that efficient 
photon number-changing processes (such as photon splitting or double Compton 
scattering) may play a key role in the formation of the spectrum. 
Indeed, we note that the photon number ratio is roughly the same as the
temperature ratio in this case, since the luminosities are approximately 
equal. As this ratio is peaked around 2 and extending from 1.5 and 4, it appears
to be qualitatively compatible with the combined effects of photon splitting
and double Compton scattering (cfr.\ Lyubarsky 2002). 

Finally, we briefly comment on the sharp turn in the hard component
distribution at around kT$\sim$10\,keV and subsequent cut-off at
$\sim$12$\div$14\,keV which, together with the $\Sigma$T$^{3}$=const relation,
are among the most distinctive properties of Figure\,\ref{fig:2bb_3D}. Among
others, these features clearly imply that no thermal component with
characteristic temperature above the cut-off is detected, though the BAT energy
interval extend up to at least 100keV. Within the discussed scenario it is
tempting to account for the above properties in terms of photon splitting
process and, in particular, its efficiency as a function of (mainly)
flux and energy. In fact, the turn at kT$\sim$10\,keV marks the drift of the
parameter distribution from the $\Sigma$T$^{3}$=const relation, identifying 
the  passage between photon number-conservation (such as
Comptonization) and photon number-changing (photon splitting) emitting
processes. As reported by TD95, the minimum temperature at which photon
splitting is efficient in maintaining a nearly Planckian spectrum for E-mode
photons, is $\simeq$10\,keV.

\subsection{The persistent emission}

The \swift\ XRT monitoring data allowed us to study the evolution of the
persistent emission of the \src\ on relatively short timescales starting from
one day before up to one an half month after the BAT event. We clearly detected
a spectral evolution in the higher energy end of the 1--10\,keV spectrum as a
function of flux, the PL component becomes flatter for larger fluxes. The 
changes in the persistent component are temporary and last for approximately a
couple of weeks, while we detected variability on daily timescales. It is also
interesting to note that the spectral results we obtained on 2006 April 8 (see
P3 interval in Table \ref{P_spectral_parameters} and the green lines in Figures
\ref{fig:spectrum_quiesc}, \ref{fig:spectrum_contplot} and
\ref{fig:spectrum_contplot2}) are consistent, to within the uncertainties,  with
those reported by Mereghetti et al.\ (2006) from a deeper ToO \XMM\ observation 
carried out on 2006 March 1. 

Both AXPs and SGRs have shown a correlation between the
X-ray flux and the spectral hardness (see, for example, Woods et al 2007; Rea et
al. 2005a); this has been explained in terms of the onset of a
``twist'' in the magnetosphere. As discussed by  Thompson et al. 2002, magnetars
may differ from standard radio pulsar because their magnetic field is
globally twisted inside the star, up to a strength of about 10 times the
external dipole and, occasionally, this field is expected to fracture the
crust and lead to a twist of the external field. The basic idea is that when
such a  twist is formed, currents flow into the magnetosphere. As the
twist angle $\Delta\phi_\mathrm{NS}$ grows, electrons provide an increasing
optical depth to resonant cyclotron scattering, leading to the build up of a
flatter photon power-law component. At the same time,
larger returning currents produce an extra heating of the star surface and 
increased X-ray flux. Therefore, both increased activity 
and glitches are expected to be associated with erratic fracturing of the crust.
Observations obtained until 2003 were consistent with a scenario in which the
twist angle was steadily increasing before the glitch epochs, culminating in
glitch(es) and period of increased timing noise, and then decreasing, leading to
a smaller flux and a softer spectrum.  The decay time of the global twist is
given by (Thompson, Lyutikov \& Kulkarni 2002)
\begin{equation}\label{tdecay}
t_{\rm decay} 
\simeq 40\,\Delta\phi_{\rm N-S}^2\,
\left({L_X\over 10^{35}~{\rm erg~s^{-1}}}\right)^{-1}\,
\left({B_{\rm pole}\over 10^{14}~{\rm G}}\right)^2\,
\left({R_{\rm NS}\over 10~{\rm km}}\right)^3\;\;\;\;\;\;{\rm  yr}, 
\end{equation}
where $\Delta\phi_{\rm N-S}$ is the net twist between
the north and south magnetic hemispheres, $B_{\rm pole}$ is the magnetic field
strength at the pole(s). By using the inferred persistent unabsorbed luminosity
of \src\ ($\sim$10$^{35}\,\ergs$), the observed time interval, of the order
of $<$1 month, over which the spectral parameters change and  return to
the pre-burst ``forest'' event, and assuming a standard size for the neutron
star surface and a $10^{14} {\rm G} <B_{\rm pole}\,{\rm G}<10^{15}$ we infer
a twist angle in the few 10$^{-3}$ to few  10$^{-2}$ radians interval, which is
similar to that inferred for the bursting AXP \ea\ ($\sim
10^{-2}$\,rad; Woods et al. 2004).

We regard as unlikely that the hardening and softening of the power--law
component (in the 1--10\,keV band) is due to a change of the flux and/or
photon index of the additional power--law component detected by {\em
INTEGRAL} in the 20-200\,keV band flux of several AXPs and SGRs
(Kuiper et al. 2006; the \src\ emission above 20\,keV was re-discovered also by
PDS on board {\em BeppoSAX}; G\"otz et al. 2006; Esposito et al. 2007). This is
implied by the finding that the spectrum of \src\ can be accounted for by only
one power--law component covering both the soft and hard X--rays (from $\sim$1
to $\sim$200\,keV) 

We monitored the pulse shape and pulse fraction of the $\sim$5.2\,s
spin period modulation through a timing analysis of the persistent component.
Large changes were detected as a function of flux (see Figure \ref{fig:fold}):
pulses appear to be less pulsed but more structured (to within the
uncertainties) and with smaller relative amplitude for increasing fluxes.
Such a flux dependence of the pulse properties was already detected in other
SGRs and AXPs and related to changes in the persistent emission component. Among
SGRs, SGR\,1806$-$20 showed a decrease in the pulsed fraction together with the
appearance of additional peaks in the pulse profile after the giant flare of the
2004 December 27 (Rea et al.\ 2005b, Woods et al. 2007).
More recently, a similar behavior was observed in the transient AXP \cxo\ in
correspondence to the onset of an outburst (Israel et al. 2007a). Given
the almost constant properties of the BB component, variations are likely due
to the PL component which is supposed to originate into the magnetosphere. In
particular, The pulse shape changes may indicate an evolution
of the geometry of the magnetosphere. Alternatively, the magnetosphere may
randomize the outcoming pulsed photons, temporarily decreasing the pulsed
fraction.

\section{Conclusions}
On 2006 March 29, after several days of burst activity, the Soft
$\gamma$-ray Repeater \src\ displayed a series of short bursts and intermediate
flares lasting for about 30\,s, during which all the \swift\ instruments were
pointed to the source. The good statistics, the fine time resolution (4\,ms)
and the wide energy coverage (0.2--100\,keV) allowed us to carry out a
detailed study of the timing and spectral properties during both the bursts
and persistent emission. Large variations were detected and their
evolution studied over a range of timescales. Although
not all the observational properties can be accounted
for by the current magnetar scenario, 
the large majority  of them are in overall good agreement with it. In
particular, we found: 

\begin{itemize} 

\item a break, around $\sim 10^{41}\,\ergs$, in the known correlation between
the
luminosities of the two blackbody components which fit well the BAT
time-resolved and BAT$+$XRT integrated spectra. Above this value the softer
blackbody shows signs of saturation, while the luminosity of the harder
blackbody still increases up to $\sim3\times10^{41}\,\ergs$; 

\item the existence of a correlation between temperature and emitting surface of
the blackbodies ($R^2 \propto kT^{-3}$), which holds through the most
luminous phases of the flares  ($L_{\rm tot}\geq 10^{41}\,\ergs$).

\end{itemize}

The above two findings together can be interpreted in terms of the
two populations of photons from the O- and E-mode polarizations modes which are
foreseen in the magnetar scenario and the regions from which they escape:
larger and colder the O-mode photon region, harder and smaller the
E-mode one. 

Moreover, we note that the maximum observed luminosity from flares ($\sim
3\times 10^{41}\,\ergs$) is attained by the harder blackbody component at an
effective temperature of $\sim$10\,keV and radius of $\sim$15\,km, which is
similar to the inferred magnetic Eddington luminosity at the same radius, for a
surface dipole field $\simeq 8 \times 10^{14}$\,G (as deduced
from the spindown of \src).

Finally, based on our analysis, we do not see any significant difference between
IFs and short bursts. Both IFs and short bursts seem to form a continuum in
terms of spectral properties (besides duration and fluence).

As far as the persistent emission properties are concerned, we found: 

\begin{itemize}

\item a clear correlation between the PL photon index $\Gamma$ and the
persistent flux, such that flatter PL components correspond to larger
fluxes. The source returned to the usual spectral shape after approximately 
one week from the burst ``forest'' event;

\item changes in the modulation of the 5.2\,s spin period, with the pulse
profiles becoming more structured (to within the uncertainties) for
increasing persistent flux. The pulse properties returned to their usual
state within one week, as well. 
\end{itemize}

The above results are not dissimilar from those observed in
other magnetar candidates, although this is the first time that variations are
detected on timescales of an hour (for the pulses) and days (for the spectra).
These findings are qualitatively in agreement with the expected properties 
of a ``twisted'' magnetosphere. 

More generally, the unprecedented high quality of the \swift\ data allowed us to
gather new important insights into the emission mechanism properties, at a
detail level which had not been attained before. We hope that findings like the
ones presented in this paper will stimulate a more detailed theoretical
treatment of the physics of SGR and AXP
bursts.

\acknowledgments

This work is partially supported at OAR through Agenzia Spaziale Italiana (ASI),
Ministero  dell'Istruzione, Universit\`a e Ricerca Scientifica e Tecnologica
(MIUR -- COFIN), and Istituto Nazionale di Astrofisica (INAF) grants.
This work is supported at OAB and OAR by ASI grants I/011/07/0,
I/088/06/0, and at OAB by MIUR PRIN 2005025417. We gratefully acknowledge the
contributions of dozens of members of the BAT, XRT, and UVOT Teams at OAB, PSU,
UL, GSFC, ASDC, and MSSL and their subcontractors,  who helped make the \swift\
mission possible. GLI would like to thank Alaa Ibrahim for useful discussions on
the comparison between the burst\ spectral properties of \sgra\ and \src, Marco
Feroci for comparison with burst properties of \src\ as observed by BeppoSAX,
Maxim Lyutikov 
for comments on the discussion section,   and Peter Woods for careful reading of
the manuscript 
and valuable suggestions for its improvement.

{\it Facilities:} \facility{Swift (XRT)}.
{\it Facilities:} \facility{Swift (BAT)}.
{\it Facilities:} \facility{Swift (UVOT)}.

\end{document}